\documentclass[10pt,conference]{IEEEtran}
\IEEEoverridecommandlockouts 
\usepackage{graphicx}
\usepackage{amsmath}
\usepackage{amssymb}
\usepackage{setspace}
\usepackage{verbatim}

\usepackage{amsthm}
\usepackage{bbm}
\usepackage{multirow}
\usepackage{psfrag}
\usepackage{booktabs}

\newtheorem{thm}{Theorem}

\newtheorem{defn}[thm]{Definition}
\newtheorem{ex}[thm]{Example}

\newcommand{\bX}{\mathbf{X}}
\newcommand{\bY}{\mathbf{Y}}
\newcommand{\bZ}{\mathbf{Z}}
\newcommand{\bW}{\mathbf{W}}

\newcommand{\bV}{\mathbf{V}}

\newcommand{\bm}{\mathbf{m}}

\newcommand{\btX}{\mathbf{\tilde{X}}}

\newcommand{\cN}{\mathcal{N}}
\newcommand{\cL}{\mathcal{L}}
\newcommand{\cE}{\mathcal{E}}

\newcommand{\Mh}{\hat{M}}

\newcommand{\eps}{\epsilon}

\newcommand{\Xit}{\tilde{\Xi}}
\newcommand{\cS}{\mathcal{S}}

\newcommand{\phs}{\mathbf{\Phi_{\mathcal{S}_m}}}

\newcommand{\mut}{\overline{\mu_t}}
\newcommand{\mutp}{\overline{{\mu}_{t}}'}

\newcommand{\muts}{\overline{\mu_{t,\cS_m}}}

\begin{document}
\title{\vspace{.25 in} Compressive Sensing Over Networks}

\author{
Soheil Feizi\\MIT\\Email: sfeizi@mit.edu \and Muriel M\'edard\\ MIT\\Email: medard@mit.edu \and Michelle Effros\\Caltech\\Email: effros@caltech.edu

\thanks{This material is based upon work under subcontract 18870740-37362-C, ITMANET project and award No. 016974-002 supported by AFSOR.}
}

\maketitle

\begin{abstract}

In this paper, we demonstrate some applications of compressive sensing over networks. We make a connection between compressive sensing and traditional information theoretic techniques in source coding and channel coding. Our results provide an explicit trade-off between the rate and the decoding complexity. The key difference of compressive sensing and traditional information theoretic approaches is at their decoding side. Although optimal decoders to recover the original signal, compressed by source coding have high complexity, the compressive sensing decoder is a linear or convex optimization. First, we investigate applications of compressive sensing on distributed compression of correlated sources. Here, by using compressive sensing, we propose a compression scheme for a family of correlated sources with a modularized decoder, providing a trade-off between the compression rate and the decoding complexity. We call this scheme \textit{Sparse Distributed Compression}. We use this compression scheme for a general multicast network with correlated sources. Here, we first decode some of the sources by a network decoding technique and then, we use a compressive sensing decoder to obtain the whole sources. Then, we investigate applications of compressive sensing on channel coding. We propose a coding scheme that combines compressive sensing and random channel coding for a high-SNR point-to-point Gaussian channel. We call this scheme \textit{Sparse Channel Coding}. We propose a modularized decoder providing a trade-off between the capacity loss and the decoding complexity. At the receiver side, first, we use a compressive sensing decoder on a noisy signal to obtain a noisy estimate of the original signal and then, we apply a traditional channel coding decoder to find the original signal. 

\end{abstract}

\section{Introduction}

Data compression has been a research topic of many scholars in past years. However, recently, the field of compressive sensing, originated in \cite{candes}, \cite{candes-noiseless} and \cite{donoho}, looked at the compression problem from another point of view. In this paper, we try to make a bridge between compressive sensing which may be viewed as a signal processing technique and both source coding and channel coding. By using this connection, we propose some non-trivial practical coding schemes that provide a trade-off between the rate and the decoding complexity.

Compressive sensing has provided a low complexity approximation to the signal reconstruction. Information theoretic has been mostly concerned with accuracy of the signal reconstruction under rate constraints. In this paper, we seek to provide new connections which use compressive sensing for traditional information theory problems such as Slepian-Wolf compression and channel coding in order to provide mechanisms to explicitly trade-off between the decoding complexity and the rate.

Some previous work investigated the connection between compressive sensing and information theory. For example, reference \cite{tarokh} studied the minimum number of noisy measurements required to recover a sparse signal by using Shannon information theory bounds. Reference \cite{bar} investigated the contained information in noisy measurements by viewing the measurement system as an information theoretic channel and using the rate distortion function. Also, reference \cite{rao-isit} studied the trade-offs between the number of measurements, the signal sparsity level, and the measurement noise level for exact support recovery of sparse signals by using an analogy between support recovery and communication over the Gaussian multiple access channel.

In this paper, we want to investigate applications of compressive sensing on source coding and channel coding. First, in Section \ref{sec:cs-source}, we consider distributed compression of correlated sources. In this section, by using compressive sensing, we propose a compression scheme for a Slepian-Wolf problem with a family of correlated sources. The proposed decoder is a modularized decoder, providing a trade-off between the compression rate and the decoding complexity. We call this scheme \textit{Sparse Distributed Compression}. Then, We use this compression scheme for a general multicast network with correlated sources. Here, we first decode some of the sources by a network decoding technique and then, we use the compressive sensing decoder to obtain the whole sources.

Then, in Section \ref{sec:cs-channel}, we investigate applications of compressive sensing on channel coding. We propose a coding scheme that combines compressive sensing and random channel coding for a high-SNR point-to-point Gaussian channel. We call this scheme \textit{Sparse Channel Coding}. We propose a modularized decoder, providing a trade-off between the capacity loss and the decoding complexity (i.e., the higher the capacity loss, the lower the decoding complexity.). The idea is to add intentionally some correlation to transmitted signals in order to decrease the decoding complexity. At the receiver side, first, we use a compressive sensing decoder on a noisy signal to obtain a noisy estimate of the original signal and then, we apply a traditional channel coding decoder to find the original signal.

\section{Technical Background and prior work}\label{sec:background}
 
In this section, we review some prior results, in both compressive sensing and distributed compression, which will be used in the rest of the paper. 

\subsection{Compressive Sensing Background}\label{subsec:cs-background}

Let $\bX\in\mathbb{R}^n$ be a signal vector. We say this signal is $k$-sparse if $k$ of its coordinates are non-zero and the rest are zero. Define $\alpha=\frac{k}{n}$ as the sparsity ratio of the signal. Suppose $\Phi$ is a $n\times n$ measurement matrix (hence, $\bY=\Phi\bX$ is the measurement vector.). If $\Phi$ is non-singular, by having $\bY$ and $\Phi$, one can recover $\bX$. However, \cite{candes-noiseless} and \cite{donoho} showed that, one can recover the original sparse signal by having far fewer measurements than $n$. Suppose for a given $n$ and $k$, $m$ satisfies the following:

\begin{equation}\label{eq:mnk_noiseless}
m \geq \rho k \log(n/k)
\end{equation}

where $\rho$ is a constant. Let $\cS$ be the power set of $\{1,2,...,m\}$, and $\cS_m$ represent a member of $\cS$ with cardinality $m$ (e.g., $\cS_m=\{1,2,...,m\}$). A $m\times n$ matrix $\phs$ is formed of rows of $\Phi$ whose indices belong to $\cS_m$. $\phs$ represents an incomplete measurement matrix. 

Reference \cite{candes-rip} showed that, if $m$ satisfies (\ref{eq:mnk_noiseless}) and $\phs$ satisfies the \textit{Restricted Isometry Property} (which will be explained later), having these $m$ measurements is sufficient to recover the original sparse vector. In fact, it has been shown in \cite{candes-rip} that, $\bX$ is the solution of the following linear programming: 

\begin{align}\label{eq:cs-opt-background}
\min\quad& \|\bX\|_{\cL_1} \\
subject\ to\quad& \bY=\phs\bX,\nonumber
\end{align}

where $\|.\|_{\cL_1}$ represents the $\cL_1$ norm of a vector. Let us refer to the complexity of this optimization by $CX_{cs}^{noiseless}(m,n)$. We say $\phs$ satisfies the restricted isometry property (RIP) iff there exists $0<\delta_k<1$ such that, for any vector $\bV\in\mathbb{R}^n$, we have:

\begin{equation}\label{eq:rip}
(1-\delta_{k})\|\bV\|_{\cL_2}^2\leq \|\phs\bV\|_{\cL_2}^2\leq (1+\delta_{k})\|\bV\|_{\cL_2}^2.
\end{equation}
 Reference \cite{candes-rip} showed if,

\begin{equation}
\delta_k+\delta_{2k}+\delta_{3k}<1,
\end{equation}

then, the optimization (\ref{eq:cs-opt-background}) can recover the original $k$-sparse signal. This condition leads to (\ref{eq:mnk_noiseless}). Intuitively, if $\phs$ satisfies the RIP condition, it preserves the Euclidean length of $k$ sparse signals. Hence, these sparse signals cannot be in the null-space of $\phs$. Moreover, any $k$ subset of columns of $\phs$ are nearly orthogonal if $\phs$ satisfies RIP. Most of matrices which satisfy RIP are random matrices. Reference \cite{bar-rip} showed a connection between compressive sensing, $n$-widths, and the Johnson-Lindenstrauss Lemma. Through this connection, \cite{database} provided an elegant way to find $\delta_k$ by using the concentration theorem for random matrices and proposed some data-base friendly matrices (binary matrices) satisfying RIP.

Now, suppose our measurements are noisy (i.e., $\bY=\phs\bX +\bZ$, where $\|\bZ\|_{\cL_2}=\eps$.). Then, reference \cite{candes} showed that if $\phs$ satisfies RIP (\ref{eq:rip}) and $\bX$ is sufficiently sparse, the following optimization recovers the original signal up to the noise level:

\begin{align}\label{eq:cs-opt-noisy-background}
\min\quad& \|\bX\|_{\cL_1} \\
subject\ to\quad& \|\bY-\phs\bX\|_{\cL_2}\leq \eps.\nonumber
\end{align}

Hence, if $\bX^{*}$ is a solution of (\ref{eq:cs-opt-noisy-background}), then $\|\bX-\bX^{*}\|_{\cL_2}\leq \beta \eps$, where $\beta$ is a constant. To have this, the original vector $\bX$ should be sufficiently sparse. Specifically, \cite{candes} showed that, if 
\begin{equation}\label{eq:mnk-noisy}
\delta_{3k}+3\delta_{4k}<2,
\end{equation}
then, the results hold. We refer to the complexity of this optimization by $CX_{cs}^{noisy}(m,n)$. 

  \begin{figure}[t]
	\centering
    \includegraphics[width=8.5cm,height=5.5cm]{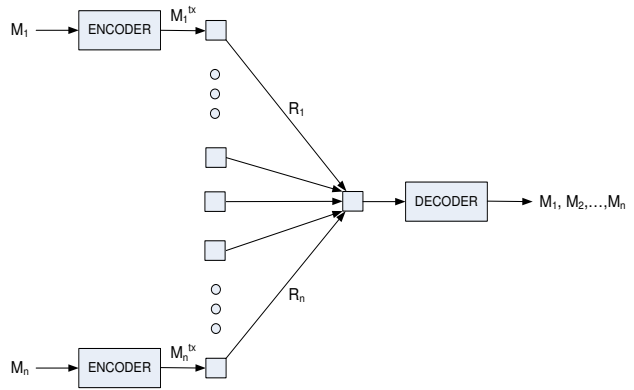}
    \caption{A depth one multicast tree with $n$ sources.}
    \label{fig:one-stage}
  \end{figure}

\subsection{Distributed Compression Background}\label{subsec:sw-background}

In this section, we briefly review some prior results in distributed compression. In Section \ref{sec:cs-source}, we will use these results along with compressive sensing results, making a bridge these two subjects. 

Consider the network depicted in Figure \ref{fig:one-stage}. This is a depth one tree network with $n$ sources. Suppose each source $i$, has messages referred by $M_i$, wishes to send to the receiver. To do this, each source $i$ sends its message with a rate $R_i$ (i.e., say $\bm_i$ has $m$ messages of source $i$. This source maps $\bm_i$ to $\{1,2,...,2^{mR_i}\}$.). Let $R=\sum_{i=1}^{n} R_i$. According to Slepian-Wolf compression (\cite{sw}), one needs to have,

\begin{equation}\label{eq:sum-rate-dep}
R \geq H(M_1,M_2,...,M_n).
\end{equation}

By using linear network coding, reference \cite{random} extended Slepian-Wolf compression for a general multicast problem. Note that, the proposed decoders in \cite{sw} and \cite{random} (a minimum entropy or a maximum probability decoder) have high complexity (\cite{kornerbook}). For a depth one tree network, references \cite{cham}, \cite{todd-medard}, \cite{ldpc-gal}, \cite{ldpc2}, \cite{turbo1} and \cite{turbo2} provide some low complexity decoding techniques for Slepian-Wolf compression. However, none of these methods can explicitly provide a trade-off between the rate and the decoding complexity. For a general multicast network, it has been shown in \cite{ram} that, there is no separation between Slepian-Wolf compression and network coding. For this network, a low complexity decoder for an optimal compression scheme has not been proposed yet.

\section{Compressive Sensing and Distributed Source Coding}\label{sec:cs-source}

In this part, we propose a distributed compression scheme for correlated sources providing a trade-off between the decoding complexity and the compression rate. References \cite{sw} and \cite{random} propose optimal distributed compression schemes for correlated sources, for depth-one trees and general networks, respectively. However, their proposed compression schemes need to use a minimum entropy or a maximum probability decoder, which has high complexity (\cite{kornerbook}). On the other hand, compressive sensing provides a low complexity decoder, by using a convex optimization, to recover a sparse vector from incomplete and contaminated observations (e.g., \cite{candes}, \cite{candes-noiseless}, \cite{donoho} and \cite{candes-rip}). Our aim is to use compressive sensing over networks to design compression schemes for correlated sources.

Note that, our proposed techniques can be applied on different network topologies, with different assumptions and constraints. Here, to illustrate the main ideas, first we consider a noiseless tree network with depth one and with correlated sources. Then, in Section \ref{subsec:cs-source-general}, we extend our techniques to a general multicast network.   

\subsection{A one-stage tree with $n$-correlated sources}\label{subsec:cs-source-one-stage}

Consider the network shown in Figure \ref{fig:one-stage}, which is a one-stage noiseless multicast tree network with $n$-sources. First, consider a case where sources are independent. Hence, the problem reduces to a classical source coding problem. Say each source is transmitting with a rate $R_i$. Denote $R=\sum_{i=1}^{n} R_i$. It is well-known that the following sum-rate for this network is the minimum required sum-rate: 

\begin{equation}\label{eq:min-cut-indep}
R \geq \sum_{i=1}^{n} H(M_i),
\end{equation}

where $M_i$ is the message random variable of source $i$. Let us refer to the complexity of its decoder at the receiver as $CX_{indep}(n)$, since sources are independent.

Now, we consider a case where sources are correlated. We formulate the correlation of sources as follows:

\begin{defn}

Suppose $\mu_{i,t}$ represents a realization of the $i^{th}$ source message random variable at time $t$ (i.e., a realization of $M_i$ at time $t$). Hence, the vector $\overline{\mu_t}=\{\mu_{1,t},...,\mu_{n,t}\}$ is the sources' message vector at time $t$. We drop the subscript $t$ when no confusion arises. Suppose sources are correlated so that there exists a transform matrix under which the sources' message vector is $k$-sparse (i.e., there exists a $n\times n$ transform matrix $\Phi$ and a $k$-sparse vector $\overline{{\mu}_{t}^{'}}$ such that $\mut=\Phi\mutp$.). Let $\cS$ be the power set of $\{1,2,...,n\}$, and $\cS_m$ denote members of $\cS$ whose cardinality are $m$. Suppose $\Phi_{\cS_m}$, defined as in Section \ref{subsec:cs-background}, satisfies RIP (\ref{eq:rip}), where $m$ satisfies (\ref{eq:mnk_noiseless}) for a given $n$ and $k$. We have, $\muts=\phs\mutp$, where $\muts$ is components of $\mut$ whose indices belong to $\cS_m$. We refer to these sources as \textit{$k$-sparsely correlated sources}.

\end{defn}

If we know zero locations of $\mutp$ and if the transform matrix is non-singular, by having each subset containing $k$ sources, the whole vector $\mut$ and $\mutp$ can be computed. Hence, by the data processing inequality, we have,

\begin{equation}
H(M_{i_1},...,M_{i_k})+n H_b(\alpha)=H(M_1,...,M_n),
\end{equation}

for and $1\leq i_1<i_2<...<i_k\leq n$ where $M_{i_j}$ is the message of source $i_j$. Also, by \cite{candes-noiseless}, having $m$ sources, where $m$ satisfies (\ref{eq:mnk_noiseless}) for a given $k$ and $n$, allows us to recover the whole vector $\mutp$ and therefore, $\mut$. Hence,

\begin{equation}\label{eq:entropy-rate-comp}
H(M_{i_1},...,M_{i_m})= H(M_1,...,M_n).
\end{equation}

\begin{ex}\label{ex:source}
For an example of $k$-sparsely correlated sources, suppose each coordinates of $\mutp$ is zero with probability $1-\alpha$, otherwise it is uniformly distributed over $\{1,2,...,2^R\}$. Suppose the entries of $\Phi$ are independent realizations of Bernoulli random variables:

\begin{eqnarray}
\Phi_{i,j}=\left\{\begin{array}{ccl}                                                       
 \frac{1}{\sqrt{m}} & & \mbox{with prob. } \frac{1}{2}\\                                        
-\frac{1}{\sqrt{m}} & & \mbox{with prob. } \frac{1}{2}
\end{array}
.\right.
\end{eqnarray}

Thus, as shown in \cite{database}, $\phs$ satisfies RIP (\ref{eq:rip}). Also, we have, 

\begin{eqnarray}
H(M_1,...,M_n)&=&n H_b(\alpha)+ k (R+1)\nonumber\\
&\approx& n H_b(\alpha)+ k R.
\end{eqnarray}

The entropy of each source random variable for large $n$ can be computed approximately as follows:

\begin{equation}
H(M_i)\approx R+\frac{1}{2}\log(k).
\end{equation}
\end{ex}

Note that, the individual entropies of $k$-sparsely correlated sources are roughly the same. Now, we want to use compressive sensing to have a distributed source coding providing a trade-off between the compression rate and the decoding complexity. 

We shall show that, if sources are $k$-sparsely correlated, if we have $m$ of them at the receiver at each time $t$, by using a linear programming, all sources can be recovered at that time. Hence, instead of sending $n$ correlated sources, one needs to transmit $m$ of them where $m<<n$. We assume that the entropies of sources are the same. Hence, each source transmits with probability $\gamma=\frac{m}{n}$. Therefore, by the law of large numbers, for large enough $n$, we have $m$ transmitting sources. For the case of non-equal source entropies, one can a priori choose $m$ sources with the lowest entropy as the transmitting sources.

In Table \ref{tab:methods}, we compare four different schemes: 

\begin{itemize}

\item The first scheme which Theorem \ref{thm:cs-sw} is about, is called \textit{Sparse Distributed Compression} with independent coding (SDCIC). The idea is to send $m$ sources from $n$ of them, assuming they are independent (i.e., their correlation is not used in coding). In this case, at the receiver, first we decode these $m$ transmitted messages and then, we use a compressive sensing decoder to recover the whole sources. The decoding complexity is $CX_{indep}(m)+CX_{cs}^{noiseless}(m,n)$. The required min-cut rate between each receiver and sources is $\sum_{j=1}^{m} H(M_{i_j})$. 

\item In the second scheme referred in Table \ref{tab:methods} as the \textit{Slepian-Wolf} coding, one performs an optimal source coding on correlated sources. The required min-cut rate between sources and the receiver is $H(M_1,...,M_n)$ which is less than $\sum_{j=1}^{m} H(M_{i_j})$. However, at the receiver, we need to have a Slepian-Wolf decoder with complexity $CX_{sw}(n)$. 

\item One can have a combination of sparse distributed compression and Slepian-Wolf compression. To do this, instead of $n$ sources, we transmit $m$ of them by using a Slepian-Wolf coding. The required min-cut rate would be $H(M_{i_1},...,M_{i_m})$ which by (\ref{eq:entropy-rate-comp}), $H(M_1,...,M_n)= H(M_{i_1},...,M_{i_m})\leq \sum_{j=1}^{m} H(M_{i_j})$. At the receiver, first we use a Slepian-Wolf decoder with complexity $CX_{sw}(m)$ to decode the $m$ transmitted sources' information. Then, we use a compressive sensing decoder to recover the whole sources. 

\item The fourth method is a naive way of transmitting $n$ correlated sources so that we have an easy decoder at the receiver. We call this scheme a \textit{Naive Correlation Ignorance} method. In this method, we simply ignore the correlation in the coding scheme. In this naive scheme, the required min-cut rate is $\sum_{i=1}^{n} H(M_i)$ with a decoding complexity equal to $CX_{indep}(n)$. Note that, in sparse distributed compression with independent coding, the required min-cut rate is much less than this scheme ($\sum_{j=1}^{m} H(M_{i_j})<<\sum_{i=1}^{n} H(M_i)$).   

\end{itemize}

\begin{table}\label{tab:methods}
\begin{center}
	\caption{Comparison of Distributed Compression Methods of Correlated Sources}\label{tab}
\resizebox{8cm}{1cm} {
  \begin{tabular}{ccc}
\toprule
\textbf{Compression Methods} & \textbf{Minimum Min-Cut Rate} & \textbf{Decoding Complexity} \\
\toprule
\multirow{1}{*}{\textbf{SDCIC}} & $\sum_{j=1}^{m} H(M_{i_j})$
 &$CX_{indep}(m)+ CX_{cs}^{noiseless}(m,n)$\\ 
\midrule
\multirow{1}{*}{\textbf{Slepian-Wolf (SW)}} & $H(M_1,...,M_n)$
 &$CX_{sw}(n)$\\
\midrule
\multirow{1}{*}{\textbf{Combination of SDC and SW}} & $H(M_1,...,M_n)$
 &$CX_{sw}(m)+CX_{cs}^{noiseless}(m,n)$\\ 
\midrule
\multirow{1}{*}{\textbf{Naive Correlation Ignorance}} & $\sum_{i=1}^{n} H(M_i)$
 &$CX_{indep}(n)$\\  
\bottomrule
\end{tabular}
}
\end{center}
\end{table}

The following theorem is about sparse distributed compression with independent coding:

\begin{thm}\label{thm:cs-sw}
For a depth one tree network with $n$-sources, $k$-sparsely correlated, the required min-cut rate between the receiver and sources in sparse distributed compression with independent coding (SDCIC) is, 
\begin{equation}\label{eq:mincut-sparse}
R\geq \sum_{j=1}^{m} H(M_{i_j}),
\end{equation}

where $m$ is the smallest number satisfies (\ref{eq:mnk_noiseless}) for a given $n$ and $k$ and $i_j$ is the index of $j^{th}$ active source. Also, the decoding complexity of SDCIC method is $CX_{indep}(m)+CX_{cs}^{noiseless}(m,n)$.  
\end{thm}

\begin{proof} 
For a given $n$ (number of sources) and $k$ (the sparsity factor), we choose $m$ to satisfy (\ref{eq:mnk_noiseless}). Now, suppose each source is transmitting its block with probability $\gamma$ where $\gamma=\frac{m}{n}$. Hence, by the law of large numbers, the probability that we have $m$ transmitting sources' blocks goes to one for sufficiently large $n$. Without loss of generality, say sources $1$, ..., $m$ are transmitting (i.e., active sources) and other sources are idle. If $R_i\geq H(M_i)$ for active sources and zero for idle sources, active sources can transmit their messages to the receiver and the required min-cut rate between the receiver and sources would be $R\geq \sum_{j=1}^{m} H(M_{i_j})$. Let $\cS_m$ indicate active sources' indices. First, at the receiver, we decode these $m$ active sources. The complexity of this decoding part is $CX_{indep}(m)$ because we did not use their correlation in the coding scheme. Hence, we have $\muts$, active sources' messages at time $t$ at the receiver (in this case, $\muts=(\mu_{1,t}, ..., \mu_{m,t})$). Then, by having $\muts$ and using the sparsity of $\mutp$, the following optimization can recover the whole sources: 

\begin{align}\label{eq:cs-opt-net}
\min\quad& \|\mutp\|_{\cL_1} \\
subject\ to\quad&  \muts=\Phi_{\cS_m} \mutp.  \nonumber 
\end{align}

The overall decoding complexity is $CX_{indep}(m)+CX_{cs}^{noiseless}(m,n)$.
\end{proof}

  \begin{figure}[t]
	\centering
    \includegraphics[width=8cm,height=5cm]{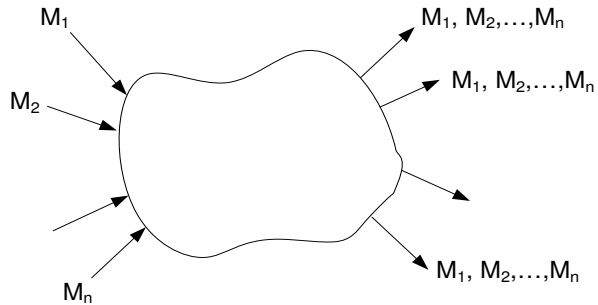}
    \caption{A general multicast network with $n$ sources.}
    \label{fig:multicast}
  \end{figure}

\subsection{A General Multicast Network with Correlated Sources}\label{subsec:cs-source-general}

In this section, we extend results of Section \ref{subsec:cs-source-one-stage} to a general noiseless multicast network. Note that, for a depth one tree network, references \cite{cham}, \cite{todd-medard}, \cite{ldpc-gal}, \cite{ldpc2}, \cite{turbo1} and \cite{turbo2} provide some low complexity decoding techniques for Slepian-Wolf compression. However, none of these methods can explicitly provide a trade-off between the rate and the decoding complexity. For a general multicast network, it has been shown in \cite{ram} that, there is no separation between Slepian-Wolf compression and network coding. For this network, a low complexity decoder for an optimal compression scheme has not been proposed yet. There are few practical approaches and they rely on small topologies \cite{multiprac}. In this section, by using compressive sensing, we propose a coding scheme which provides a trade-off between the compression rate and the decoding complexity.

Consider a multicast network shown in Figure \ref{fig:multicast} which has $n$ sources and $T$ receivers. If sources are independent, reference \cite{ahl2000} showed that the minimum required min-cut rate between each receiver and sources is as (\ref{eq:min-cut-indep}). This rate can be achieved by using linear network coding over the network. Reference \cite{random} showed that random linear network coding can perform arbitrarily closely to this rate bound. Say $CX_{indep}^{nc}(n)$ is the complexity of its decoder.

Now, consider a case when sources are correlated. Reference \cite{random} extended Slepian-Wolf compression result (\cite{sw}) to a multicast network by using network coding. The proposed decoder is a minimum entropy or a maximum probability decoder whose complexity is high (\cite{kornerbook}). We refer to its complexity by $CX_{sw}^{nc}(n)$.

In this section, we consider sources to be $k$-sparsely correlated. We want to use compressive sensing along with network coding to propose a distributed compression scheme providing a trade-off between the compression rate and the decoding complexity.

In the following, we compare four distributed compression methods for a general multicast network:

\begin{itemize}

\item In the first scheme, we use sparse distributed compression with independent coding. With high probability, we have $m$ active sources. Then, we perform network coding on these active sources over the network, assuming they are independent. At the receiver, first we decode these $m$ active sources and then, by a compressive sensing decoder, we recover the whole sources. The decoding complexity is $CX_{indep}^{nc}(m)+CX_{cs}^{noiseless}(m,n)$. The required min-cut rate between each receiver and sources is $\sum_{j=1}^{m} H(M_{i_j})$. 

\item In the second scheme, we use an extended version of Slepian-Wolf compression for a multicast network (\cite{medard2003}). Reference \cite{medard2003} considers a vector linear network code that operates on blocks of bits. The required min-cut rate between sources and the receiver is $H(M_1,...,M_n)$. At the receiver, one needs to have a minimum entropy or a maximum probability decoder with complexity $CX_{sw}^{nc}(n)$. 

\item A combination of sparse distributed compression and Slepian-Wolf compression can provide a trade-off between the compression rate and the decoding complexity. For example, instead of $n$ sources, one can transmit $m$ of them by using a Slepian-Wolf coding for multicast networks. The required min-cut rate would be $H(M_{i_1},...,M_{i_m})$. At the receiver, first we use a Slepian-Wolf decoder with complexity $CX_{sw}^{nc}(m)$ to decode $m$ active sources. Then, we use a compressive sensing decoder to recover the whole sources. 

\item In a naive correlation ignorance method, we simply ignore the correlation in the coding scheme. In this naive scheme, the required min-cut rate is $\sum_{i=1}^{n} H(M_i)$ with a decoding complexity equal to $CX_{indep}^{nc}(n)$. 
\end{itemize}

The following theorem is about sparse distributed compression with independent coding for a multicast network:

\begin{thm}\label{thm:cs-multicast}
For a general multicast network with $n$-sources, $k$-sparsely correlated, the required min-cut rate between each receiver and sources in sparse distributed compression with independent coding is, 
\begin{equation}\label{eq:mincut-sparse}
R\geq \sum_{j=1}^{m} H(M_{i_j}),
\end{equation}

where $m$ is the smallest number satisfies (\ref{eq:mnk_noiseless}) for a given $n$ and $k$. Also, the decoding complexity is $CX_{indep}^{nc}(m)+CX_{cs}^{noiseless}(m,n)$.  
\end{thm} 

\begin{proof}
The proof is similar to the one of Theorem \ref{thm:cs-sw}. The only difference is that, here, one needs to perform linear network coding on active sources over the network without using the correlation among them. At the receiver, first, these $m$ active sources' blocks are decoded (the decoding complexity of this part is $CX_{indep}^{nc}(m)$). Then, a compressive sensing decoder (\ref{eq:cs-opt-net}) is used to recover the whole sources.
\end{proof}

  \begin{figure}[t]
	\centering
    \includegraphics[width=8cm,height=4.5cm]{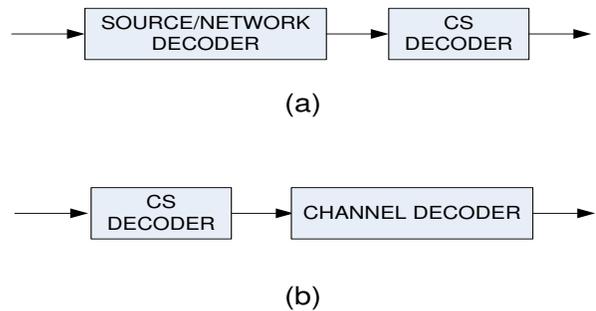}
    \caption{A modularized decoder for a) compressive sensing with source coding, (b) compressive sensing with channel coding.}
    \label{fig:comp}
  \end{figure}

\textit{Remark:} In this section, we explained how compressive sensing can be used with distributed source coding. At the receiver side, first, we decode the active sources by a network decoding technique and then, we use a compressive sensing decoder to recover the whole sources. This high level modularized decoding scheme of sparse distributed compression is depicted in Figure \ref{fig:comp}-a.  On the other hand, when we use compressive sensing for channel coding (which will be explained in detail in Section \ref{sec:cs-channel}), we switch these two decoding modules (as shown in Figure \ref{fig:comp}-b). In other words, first, we use a compressive sensing decoder on a noisy signal to obtain a noisy estimate of the original signal and then, we use a channel decoder to find the original message from this noisy estimate.

\section{Compressive Sensing and Channel Coding}\label{sec:cs-channel}

\begin{figure}[t]
	\centering
    \includegraphics[width=8cm,height=2cm]{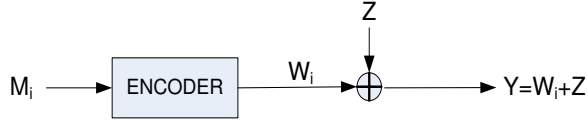}
    \caption{A Point-to-Point Gaussian Channel}
    \label{fig:ptp}
  \end{figure}

In this section, we make a bridge between compressive sensing, which is more a signal processing technique, and channel coding. We consider a high-SNR Gaussian point-to-point channel depicted in Figure \ref{fig:ptp}. We propose a coding scheme that combines compressive sensing and random channel coding with a modularized decoder to provide a trade-off between the capacity loss and the decoding complexity (i.e., the higher the capacity loss, the lower the decoding complexity.). We call this scheme \textit{Sparse Channel Coding}. We add intentionally some correlation to transmitted signals to decrease the decoding complexity. This is an example of how compressive sensing can be used with channel coding and could be extended to other types of channels.

Consider a point to point channel with additive Gaussian noise with noise power $N$ and transmission power constraint $P$, shown in Figure \ref{fig:ptp}. Suppose we are in a high SNR regime (i.e., $\frac{P}{N}\gg 1$). The capacity of this channel is,

\begin{equation}\label{eq:capacity}
C\approx\frac{1}{2}\log{(\frac{P}{N})}.
\end{equation}

An encoding and a decoding scheme for this channel can be found in \cite{kornerbook}. As explained in \cite{kornerbook}, at the receiver, one needs to use a $m$-dimensional maximum likelihood decoder where $m$, the code length, is arbitrarily large. We refer to the complexity of such a decoder as $CX_{ml}(m)$. Our aim is to design a coding scheme to have a trade-off between the rate and the decoding complexity.

Suppose $m$ is arbitrarily large. Choose $n$ and $k$ to satisfy (\ref{eq:mnk-noisy}).

\begin{defn}\label{def:point-to-point}
A $(2^{mR},m)$ sparse channel code for a point-to-point Gaussian channel with power constraint $P$  satisfies the following:
\begin{itemize}
\item Encoding process: message $i$ from the set $\{1,2,...,2^{mR}\}$ is assigned to a $m$-length vector $\bW_i$, satisfying the power constraint; that is, for each codeword, we have,

\begin{equation}
\frac{1}{m}\sum_{j=1}^{m} W_{ij}^2 \leq P,
\end{equation}

where $W_{ij}$ is the $j^{th}$ coordinate of codeword $\bW_i$.
 
\item Decoding process: the receiver receives $\bY$, a noisy version of the transmitted codeword $\bW_i$ (i.e., $\bY=\bW_i+\bZ$, where $\bZ$ is a Gaussian noise vector with i.i.d. coordinates with power $N$). A decoding function $g$ maps $\bY$ to the set $\{1,2,...,2^{mR}\}$. The decoding complexity is $CX_{ml}(k)+CX_{cs}^{noisy}(m,n)$.
\end{itemize}

A rate $R$ is achievable if $P_e^m \to 0$ as $m\to\infty$, where,

\begin{equation}
P_e^{m}=\frac{1}{2^{mR}}\sum_{i=1}^{2^{mR}} Pr(g(\bY)\neq i|M=i),
\end{equation}

and $M$ represents the transmitted message. 

\end{defn}

\begin{thm}\label{thm:scc}
For a high-SNR Gaussian point-to-point channel described in Definition \ref{def:point-to-point}, the following rate is achievable, while the decoding complexity is $CX_{ml}(k)+CX_{cs}^{noisy}(m,n)$:

\begin{equation}\label{eq:ptp-rate}
R=\frac{k}{m} C+\frac{n}{m}H_b(\alpha)+\frac{k}{2m}\log(\frac{1}{\beta(1+\delta_k)}),
\end{equation}

where $C$ is the channel capacity (\ref{eq:capacity}), $\alpha$, $\beta$ and $\delta_k$ are parameters defined in Section \ref{subsec:cs-background}, and $H_b(.)$ is the binary entropy function.  
\end{thm}

Before expressing the proof, let us make some remarks:

\begin{itemize}
\item if we have (\ref{eq:mnk_noiseless}) (i.e., $m/n=\rho \alpha \log(1/\alpha)$), the achievable rate can be approximated as follows:

\begin{equation}\label{eq:ptp-rate-apx}
R\approx\frac{1}{\log(1/\alpha)} C + \frac{1}{\alpha \log(1/\alpha)} H_b(\alpha)
\end{equation}

where $\alpha=k/n$ is the sparsity ratio. The first term of (\ref{eq:ptp-rate-apx}) shows that the capacity loss factor is a $\log$ function of the sparsity ratio. The term $\frac{1}{\alpha \log(1/\alpha)} H_b(\alpha)$ is the rate gain that we obtain by using compressive sensing in our scheme. Note that, since the SNR is sufficiently high, the overall rate would be less than the capacity. 

\item The complexity term $CX_{ml}(k)$ is an exponential function, while $CX_{cs}^{noisy}(m,n)$ is a polynomial.  

\end{itemize}

\begin{proof}
We use random coding and compressive sensing arguments to propose a concatenated channel coding scheme satisfying ($\ref{eq:ptp-rate}$). First, we explain the encoding part:

\begin{itemize}
\item Each message $i$ is mapped randomly to a $k$-sparse vector $\bX_i$ with length $n$ so that its non-zero coordinates are i.i.d. drawn from $\cN(0,\frac{m}{k(1+\delta_k)}P)$. $n$ is chosen to satisfy (\ref{eq:mnk-noisy}), for a given $k$ and $m$. 
\item To obtain $\bW_i$, we multiply $\bX_i$ by a $m\times n$ matrix ($\phs$) satisfying the RIP condition (\ref{eq:rip}) (i.e., $\bW_i=\phs\bX_i$). We send $\bW_i$ through the channel.
\end{itemize}

In fact, it is a concatenated channel coding scheme \cite{forney}. The outer layer of this coding is the sparsity pattern of $\bX_i$. The inner layer of this coding is based on random coding around each sparisty pattern. To satisfy transmission power constraint $P$, we generate each non-zero coordinate of $\bX_i$ by $\cN(0,\frac{m}{k(1+\delta_k)}P)$. Note that, the encoding matrix $\phs$ is known to both encoding and decoding sides.

At the receiver, we have a noisy version of the transmitted signal. Suppose message $i$ has been transmitted. Hence, $\bY=\bW_i+\bZ=\phs\bX_i+\bZ$. The decoding part is as follows:
\begin{itemize}
\item Since $\bX_i$ is $k$-sparse, first, we use a compressive sensing decoder to find $\btX_i$ such that $\|\bX_i-\btX_i\|_{\cL_2}^2\leq \beta m N$, as follows, where $\beta$ is the parameter of (\ref{eq:cs-opt-noisy-background}):

\begin{align}\label{eq:cs-opt}
\min\quad& \|\bX\|_{\cL_1} \\
subject\ to\quad& \frac{1}{m}\|\bY-\phs\bX_i\|_{\cL_2}^2\leq N.\nonumber
\end{align}

We assume that the complexity of this convex optimization is smaller than the one of a maximum likelihood decoder.

\item Since we are in a high-SNR regime, by having $\btX_i$, we can find the sparsity pattern of $\bX_i$. It gives us the outer layer code of message $i$. The higher the outer layer code rate, the higher the required SNR. We shall develop this argument with more detail later (\ref{eq:outerlayer-error}).

\item Having the sparsity pattern of $\bX_i$, we use a maximum likelihood decoder in a $k$-dimensional space (i.e., non-zero coordinates of $\bX_i$) to find non-zero coordinates of $\bX_i$ (the inner layer code). The complexity of this decoder is denoted by $CX_{ml}(k)$.
\end{itemize}

Before presenting the error probability analysis, let us discuss different terms of (\ref{eq:ptp-rate}). Since in our coding scheme, for each message $i$, we send a $k$-sparse signal by sending a $m$-length vector, we have the fraction $\frac{k}{m}$ before the capacity term $C$. The capacity term $C$ in (\ref{eq:ptp-rate}) comes from the inner layer coding scheme. The term $\frac{n}{m}H_b(\alpha)$ comes from the outer layer coding scheme. Note that, since $\alpha$ is a small number, $H_b(\alpha)$ is close to one. Also, the ratio $\frac{n}{m}$ depends on the outer layer code rate. Here, we assumed the SNR is sufficiently high so that we can use all possible outer layer codes. The term $\frac{k}{2m}\log(\frac{1}{\beta(1+\delta_k)})$ is because of the power constraint and the RIP condition. Note that, if one performs a time-sharing based channel coding, a rate $\frac{k}{m} C$ is achievable with a decoding complexity $CX_{ml}(k)$. By using the compressive sensing, we obtain additional rate terms because of the outer layer coding with approximately the same complexity.

We proceed by the error probability analysis. Define $\Xi_{i}$ as the sparsity pattern of $\bX_i$ (i.e., $\Xi_{ij}=0$ if $X_{ij}=0$, otherwise, $\Xi_{ij}=1$.). Also, define $\Xit_i$ as the decoded sparsity pattern of message $i$. Without loss of generality, assume that message $1$ was transmitted. Thus, $\bY=\phs\bX_1+\bZ$. Define the following events: 

\begin{eqnarray}
&&E_0=\{\frac{1}{m}\sum_{j=1}^{m} W_{1j}^2 >P \}\nonumber\\
&&E_i=\{\frac{1}{n}\|\bX_i-\btX_1\|_{\cL_2}^2\leq \frac{\beta m}{n}N\}\nonumber\\
&&E_{p_1}=\{\Xit_1\neq \Xi_1\}.
\end{eqnarray}

Hence, an error occurs when $E_0$ occurs (i.e., the power constraint is violated), or $E_{p_1}$ occurs (i.e., the outer layer code is wrongly decoded), or $E_1^c$ occurs (i.e., the underlying sparse signal of the transmitted message $\bX_1$ and its decoded noisy version $\btX_1$ are in a distance greater than the noise level), or one of $E_i$ occurs while $i\neq 1$. Say $M$ is the transmitted message (in this case $M=1$) and $\Mh$ is the decoded message. Let $\cE$ denote the event $M\neq \Mh$. Hence,

\begin{eqnarray}\label{eq:error1}
Pr(\cE|M=1)&=&Pr(\cE)\nonumber\\
&=&Pr(E_0\bigcup E_{p_1} \bigcup E_1^c \bigcup_{j=2}^{2^{mR}} E_j )\\
&\leq& Pr(E_0)+Pr(E_{p_1})+Pr(E_1^c)\nonumber\\
&+&\sum_{j=2}^{2^{mR}} Pr(E_j),\nonumber
\end{eqnarray}

by union bounds of probabilities. We bound these probabilities term by term as follows:

\begin{itemize}
\item By the law of large numbers, we have $Pr(\|\bX_1\|_{\cL_2}^{2}>\frac{m}{1+\delta_k}P+\eps)\to 0$ as $n\to\infty$. By using the RIP condition (\ref{eq:rip}) and with probability one, we have,

\begin{equation}
\|\phs\bX_1\|_{\cL_2}^2\leq (1+\delta_k)\|\bX_1\|_{\cL_2}^2\leq mP.
\end{equation}

Hence, $Pr(E_0)\to 0$ as $n\to\infty$.

\item After using the compressive sensing decoder mentioned in (\ref{eq:cs-opt}), we have $\btX_1$ such that $\frac{1}{n}\|\bX_1-\btX_1\|_{\cL_2}^2\leq \frac{\beta m}{n} N$. Suppose this error is uniformly distributed over different coordinates. We use a threshold comparison to determine the sparsity pattern $\Xit_1$. The probability of error in determining $\Xit_1$ determines how high the SNR should be. Intuitively, if we use all possible outer layer codes (all possible sparsity patterns), we should not make any error in determining the sparsity pattern of each coordinate $j$ (i.e., $\Xit_{1j}$). Hence, we need sufficiently high SNR for a given $n$. On the other hand, if we decrease the outer layer code rate, the required SNR would be lower than the case before. Here, to illustrate how the required SNR can be computed, we assume that we use all possible sparsity patterns. Say $E_{p_{j1}}$ is the event of making an error in determining whether the $j^{th}$ coordinate is zero or not. We say $\Xit_{1j}=0$ if $X_{ij}<\tau$. Otherwise, $\Xit_{1j}=1$. Hence, for a given threshold $\tau$, we have,

\begin{eqnarray}\label{eq:outerlayer-error}
Pr(E_{p_{j1}}|\Xi_{1j}=0)&=&\phi (\frac{\tau}{\sqrt{\frac{\beta m}{n}N}})\nonumber\\
Pr(E_{p_{j1}}|\Xi_{1j}=1)&=& 1-2\phi(\frac{\tau}{\sqrt{P+\frac{\beta m}{n}N}})\nonumber
\end{eqnarray}

where $\phi(.)$ is the cumulative distribution function of a normal random variable. Hence,

\begin{eqnarray}
Pr(E_{p_1})&=&1-Pr(E_{p_1}^c)\\
&=&1-\big(1-\phi (\frac{\tau}{\sqrt{\frac{\beta m}{n}N}})\big)^{n(1-\alpha)}\nonumber\\
&\times&\big(2\phi(\frac{\tau}{\sqrt{P+\frac{\beta m}{n}N}})\big)^{n\alpha}\nonumber
\end{eqnarray}

Note that, one can choose $\tau$ and $\frac{P}{N}$ large enough to have $Pr(E_{p_1})$ arbitrarily small.
 
\item By \cite{candes-noiseless}, $Pr(E_1^c)$ is zero.
\item At the last step, we need to bound $Pr(E_j)$ for $j\neq 1$. We have,

\begin{eqnarray}\label{eq:error-bound1}
Pr(E_j)&=&Pr(\Xi_j=\Xi_1) Pr(E_j|\Xi_j=\Xi_1)\nonumber\\
&+&Pr(\Xi_j\neq \Xi_1) Pr(E_j|\Xi_j\neq \Xi_1)\nonumber\\
&=&\alpha^k (1-\alpha)^{n-k} Pr(E_j|\Xi_j=\Xi_1).
\end{eqnarray}

Note that, $Pr(E_j|\Xi_j\neq \Xi_1)$ goes to zero in a high SNR regime. By bounding the noise power of non-zero coordinates by $\beta m N$ (the whole noise power) and using random coding argument in a $k$ dimensional space, we have,

\begin{equation}\label{eq:error-bound2}
Pr(E_j|\Xi_j=\Xi_1)\dot{\leq}\frac{1}{(1+\frac{1}{\beta (1+\delta_k)}\frac{P}{N})},
\end{equation}

where $\dot{\leq}$ indicates the right hand side is less or equal than the left one in an exponential rate. Therefore, by (\ref{eq:error-bound1}) and (\ref{eq:error-bound2}), and doing some manipulations, we have, 

\begin{equation}\label{eq:error2}
P(E_j)\dot{\leq} 2^{-n(H_b(\alpha)+\frac{\alpha}{2}\log(\frac{1}{\beta (1+\delta_k)}\frac{P}{N}))},
\end{equation}
for $j=2,...,2^{mR}$.
\end{itemize}

Thus, in a sufficiently high SNR regime, by (\ref{eq:error1}) and (\ref{eq:error2}), we have,  

\begin{eqnarray}
P_e^m=Pr(\cE)&=&Pr(\cE|M=1)\\
&\dot{\leq}& 2^{mR} 2^{-n(H_b(\alpha)+\frac{\alpha}{2}\log(\frac{1}{\beta (1+\delta_k)}\frac{P}{N}))}\nonumber\\
&\dot{\leq}& 2^{m(R -\frac{n}{m}H_b(\alpha)-\frac{k}{m} C-\frac{k}{2m}\log(\frac{1}{\beta (1+\delta_k)}))}.\nonumber
\end{eqnarray}

For any given $\eps>0$, we can choose $n$ large enough to have $R\leq \frac{k}{m} C+ \frac{n}{m}H_b(\alpha)+\frac{k}{2m}\log(\frac{1}{\beta (1+\delta_k)})-\eps$ achievable.

\end{proof}

\section{Conclusions}

In this paper, we demonstrated some applications of compressive sensing over networks. We made a connection between compressive sensing and traditional information theoretic techniques in source coding and channel coding in order to provide mechanisms to explicitly trade-off between the decoding complexity and the rate. Although optimal decoders to recover the original signal compressed by source coding have high complexity, the compressive sensing decoder is a linear or convex optimization. First, we investigated applications of compressive sensing on distributed compression of correlated sources. For a depth one tree network, references (\cite{cham}-\cite{turbo2}) provide some low complexity decoding techniques for Slepian-Wolf compression. However, none of these methods can explicitly provide a trade-off between the rate and the decoding complexity. For a general multicast network, reference \cite{random} extended Slepian-Wolf compression for a multicast problem. It is showed in \cite{ram} that, there is no separation between Slepian-Wolf compression and network coding. For this network, a low complexity decoder for an optimal compression scheme has not been proposed yet. Here, by using compressive sensing, we proposed a compression scheme for a family of correlated sources with a modularized decoder providing a trade-off between the compression rate and the decoding complexity. We called this scheme \textit{Sparse Distributed Compression}. We used this compression scheme for a general multicast network with correlated sources. Here, we first decoded some of the sources by a network decoding technique and then, we used a compressive sensing decoder to obtain the whole sources.

Next, we investigated applications of compressive sensing on channel coding. We proposed a coding scheme that combines compressive sensing and random channel coding for a high-SNR point-to-point Gaussian channel. We called this scheme \textit{Sparse Channel Coding}. Our coding scheme provides a modularized decoder to have a trade-off between the capacity loss and the decoding complexity. The idea is to add intentionally some correlation to transmitted signals in order to decrease the decoding complexity. At the decoder side, first we used a compressive sensing decoder to get an estimate of the original sparse signal, and then we used a channel coding decoder in the subspace of non-zero coordinates.


\IEEEpeerreviewmaketitle



\end{document}